\documentclass[12pt, twoside]{article}
\usepackage{amsmath,amsthm}
\usepackage{amssymb}
\usepackage{enumerate}
\usepackage{graphicx}
\usepackage{enumitem}
\usepackage{braket,mleftright}

\def\barr{\begin{array}}
\def\earr{\end{array}}

\newtheorem{theorem}{Theorem}[section]
\newtheorem{lemma}[theorem]{Lemma}

\newtheorem{corollary}[theorem]{Corollary}
\theoremstyle{definition}
\newtheorem{definition}[theorem]{Definition}

\theoremstyle{remark}

\numberwithin{equation}{section}
\def\zC{\mathbb C}    
\def\zR{\mathbb R}

\def\t{\mbox{tr}\,}
\DeclareTextFontCommand{\textroman}{\fontlibertine}

\numberwithin{equation}{section}

\begin{document}
\baselineskip=17pt
\title{\textbf{Some applications of matrix inequalities in R{\'e}nyi entropy}}
\author{Hadi Reisizadeh$^{1,*}$ and S. Mahmoud Manjegani$^{2,\dagger}$\\
\textit{$^{1}$Department of Electrical and Computer Engineering,}\\
\textit{$^{2}$ Department of Mathematical Sciences,}\\
\textit{Isfahan University of Technology}\\
}
 
\date{}
\maketitle
\renewcommand{\thefootnote}{}

\footnote{$^*$hadi.reisizadeh@gmail.com}

\footnote{$^\dagger$manjgani@cc.iut.ac.ir}

\renewcommand{\thefootnote}{\arabic{footnote}}
\setcounter{footnote}{0}
\begin{abstract}
The R{\'e}nyi entropy is one of the important information measures that generalizes Shannon's entropy. The quantum R{\'e}nyi entropy has a fundamental role in quantum information theory, therefore, bounding this quantity is of vital importance. Another important quantity is R{\'e}nyi relative entropy on which R{\'e}nyi generalization of the conditional entropy, and mutual information are defined based. Thus, finding lower bound for R{\'e}nyi relative entropy is our goal in this paper. We use matrix inequalities to prove new bounds on the entropy of type $\beta$, R{\'e}nyi entropy.
\end{abstract}
\section{Introduction}
There are several entropic quantities belonging to the family of $\beta$-entropies entropies that have been shown to be useful. The entropies of type $\beta$ are defined by means of information functions \cite{ZD}. Recall that a real function $f$ defined on $[0,1]$ is an information function if it satisfies the boundary conditions
\begin{equation}\label{E1}
f(0)=f(1);~~~~ f(\frac{1}{2})=1, 
\end{equation}
and the functional equation 
\[
f(x)+(1-x)f(\frac{y}{1-x})=f(y)+(1-y)f(\frac{x}{1-y}),
\]
for all $(x,y)\in D$, where
\[
 D=\left\{ (x,y);~0\le x\le 1, 0\le y\le 1,  \mbox{and}~ x+y\le 1\right  \}.
 \]

Let $f$ be an information function and $(p_1, p_2, \cdots ,p_n)$ be an finite discrete probability distribution. Then in \cite{ZD} the entropy of the distribution $(p_1, p_2, \cdots ,p_n)$ with respect to $f$ is defined by
\[
H_n^f(p_1, p_2, \cdots ,p_n)=\sum_{i=2}^ns_if(\frac{p_i}{s_i}),~~~~ s_i=p_1+  \cdots +p_i ; ~~i=2,\cdots , n\,.
\]
\begin{definition}\cite{ZD} Let $\beta$ be a positive number. We call the real function defined in $[0,1]$ an {\tt information function of type $\beta$} if it satisfies the boundary condition (\ref{E1}) and the function equation
\[
f(x)+(1-x)^{\beta}f(\frac{y}{1-x})=f(y)+(1-y)^{\beta}f(\frac{x}{1-y}),
\]
for all $(x,y)\in D$.
\end{definition}
The {\tt entropy of type $\beta$} of a probability distribution $(p_1, p_2, \cdots ,p_n)$ is defined by 
\[
H_n^{\beta}(p_1, p_2, \cdots ,p_n)=\sum_{i=2}^ns_i^{\beta}f(\frac{p_i}{s_i}),~~~~ s_i=p_1+  \cdots +p_i ; ~~i=2,\cdots , n\,.
\]
where $f$ is an information function of type $\beta$\,.
\begin{theorem}\cite{ZD} Let $f$ be an information function of type $\beta$ with $\beta\ne 1$. Then
\[
f(x)=(2^{1-\beta}-1)^{-1}[x^{\beta}+(1-x)^{\beta}-1], \mbox{ for all} ~x\in [0,1]\,.
\]

\end{theorem}

\begin{theorem}\cite{ZD} Let $\beta$ be a positive number $\beta$ with $\beta\ne 1$. Then we have for the entropy of type $\beta$ of a probability distribution $(p_1, p_2, \cdots ,p_n)$,
\begin{equation}\label{E2}
H_n^{\beta}(p_1, p_2, \cdots ,p_n)=(2^{1-\beta}-1)^{-1}\left(\sum_{i=1}^np_i^{\beta}-1\right)\,.
\end{equation}
\end{theorem}
The Shannon's entropy $H_n(p_1, p_2, \cdots ,p_n)$ is the limit function of $H_n^{\beta}(p_1, p_2, \cdots ,p_n)$, when $\beta\rightarrow 1$.

 For a positive number $\beta$ with $\beta\ne 1$, R{\'e}nyi \cite{AR} has extended the concept of Shannon's entropy by defining the {\tt entropy of order $\beta$} of a probability distribution $(p_1, p_2, \cdots ,p_n)$ as 
\begin{equation}\label{E3}
{_\beta}H_n(p_1,p_2,...,p_n) = (1-\beta)^{-1}\log_2\sum_{i=1}^{n}p_i^\beta \,.
\end{equation}
 

Let $H$ be a  complex Hilbert space with dimension $n$. The set of linear operators  on $H$ is denoted by $M_n(\zC)$, where $M_n(\zC)$ is the set (the $C^*$-algebra) of
all $n$ by $n$ complex matrices. The conjugate transpose of matrix $A\in M_n(\zC)$ is denoted by $A^*$ or $A^\dagger$. If $A^\dagger=A $,  then $A$ is called Hermitian matrix. A Hermitian or self adjoint matrix $A\in M_n(\zC)$ is called positive semi definite (resp. positive definite) if $\langle Ax,x\rangle\geq 0$ (resp. $\langle Ax,x\rangle >0$) for each $x\in \zC^n$. The set $M_n^+(\zC)$ of all positive semi definite matrices is then a closed convex cone in $M_n(\zC)$ and makes the set of all Hermitian matrices partially ordered: for Hermitian  matrices $A$ and $B$, $A\leq B$  if and only if $B-A\in M_n^+(\zC)$ \cite{BH}.\\

\begin{lemma}\label{L1}
If $ A \ge 0 $, then for all $ r \in \mathbb{R} ; A^r \ge 0. $
\end{lemma}
\noindent {\it Proof.} It follows from definition of positive semidefinite matrix.
\qed
\begin{lemma}\label{L2}
If $A$ and $B$ are positive semidefinite matrices, then
\[ 0 \leq \t(AB) \leq \t(A)\t(B)\,. \]
\end{lemma}
\noindent {\it Proof.} This is an immediate result of the Von Neumann's trace inequality \cite{LM}.
\qed
\begin{lemma}\label{L3}
Let $A$ and $B$ be $n$ by $n$ positive semidefinite matrices, then \\
\begin{equation}\label{EE1}
 n(\det A \det B)^\frac{1}{n} \leq \t(AB).
 \end{equation}
Equality holds if and only if $ B^\frac{1}{2}AB^\frac{1}{2} = c \mathbb{I} $ ; $c \in \mathbb R^+ $.
\end{lemma}
\noindent {\it Proof.} $B$ is a positive semidefinite matrix, therefore it has a unique square root. Also, for square matrices $X$ and $Y$, the trace of $XY$ is equal to the trace of $YX$. It follows that the inequality (\ref{EE1}) can be rewritten as
\[
 (\det (B^{\frac{1}{2}}AB^{\frac{1}{2}}))^\frac{1}{n} \leq \frac{\t(B^{\frac{1}{2}}AB^{\frac{1}{2}})}{n}\,.
\]
Assuming that $\lambda_1, \lambda_2,
\cdots, \lambda_n\in \zR_0^+$ are the eigenvalues of $B^{\frac{1}{2}}AB^{\frac{1}{2}}$, by arithmetic-geometric mean inequality we have
\[
 (\det (B^{\frac{1}{2}}AB^{\frac{1}{2}}))^\frac{1}{n}=\sqrt[n]{\lambda_1 \lambda_2
\cdots \lambda_n}\le \frac{\sum_{i=1}^n\lambda_i}{n}= \frac{\t(B^{\frac{1}{2}}AB^{\frac{1}{2}})}{n}\,.
\]
Moreover, equality holds in the arithmetic-geometric mean inequality if and only if $\lambda_1= \lambda_2=
\cdots=\lambda_n$ which implies equality holds in (\ref{EE1}), if $B^{\frac{1}{2}}AB^{\frac{1}{2}}=\lambda_1 \mathbb{I}$.
It is clear that  $ B^\frac{1}{2}AB^\frac{1}{2} = c \mathbb{I} $ for some  $c \in \mathbb R^+ $, then $ n(\det A \det B)^\frac{1}{n} = \t(AB)$.
\qed 
\begin{lemma}\label{L4}
If $A$ is a positive definite matrix, then
\begin{equation}\label{EE2}
 \t(\mathbb{I}- A^{-1}) \leq \log\det(A) \leq \t(A - \mathbb{I}) \,.
\end{equation}
Equality holds if and only if $ A = \mathbb{I} $.
\end{lemma}
\noindent {\it Proof.} Let  $\lambda_1, \lambda_2,
\cdots, \lambda_n\in \zR_0^+$ be the eigenvalues of $A$. Then by functional calculus theorem, the inequality (\ref{EE2}) can be rewritten as
\[
\sum_{i=1}^n(1-\frac{1}{\lambda_i})\le \sum_{i=1}^n\log(\lambda_i)\le \sum_{i=1}^n(\lambda_i-1),
\]
or, by setting $\lambda_i=e^{u_i}$,
\[
\sum_{i=1}^n(1-e^{-u_i})\le \sum_{i=1}^nu_i\le \sum_{i=1}^n(e^{u_i}-1),
\] 
that follows from the convexity of the exponential function. Equality holds if and only if $u_i=0$ for every $1\le i\le n$ implying $\lambda_i=1$  for every $1\le i\le n$. Thus, equality holds in (\ref{EE2}) if and only if $ A = \mathbb{I} $.
\qed
\begin{definition}
An operator $\rho$ on a finite dimensional Hilbert space $H$, is called  density operator if it satisfies the following three requirements:\\
 \begin{enumerate}[label=(\roman*)]
\item $\rho$ is Hermitian,
\item $\t{\rho}$  = 1, and 
\item  $\rho$ is a positive semi-definite operator.
 \end{enumerate}
\end{definition}
 
\section{Entropy of type $\beta$ and R{\'e}nyi entropy}

First, we prove some results about the R{\'e}nyi entropy of order $\beta$.
For $n\ge 1$, we denote 
\[
\Delta_n = \left\{(p_1,p_2,...,p_n) : p_i\geq 0 , ~~  \sum_{i=1}^{n}p_i = 1 \right\}.
\]

\begin{theorem}\label{T1}
For all $\textup{(}p_1,p_2,...,p_n\textup{)} \in\Delta_n $ and positive real number $\beta$,
\begin{enumerate}
\item if $0<\beta<1$, then
\[
{_\beta}H_n(p_1,p_2,...,p_n)\geq (1-\beta)^{-1}\Big(\log_2(n - n_0) + \frac{\beta}{n - n_0}\sum_{i=1}^{n - n_0}\log_2{p_i}'\Big).
\]
\item if $\beta>1$, then
\[
{_\beta}H_n(p_1,p_2,...,p_n)\leq (1-\beta)^{-1}\Big(\log_2(n - n_0) + \frac{\beta}{n - n_0}\sum_{i=1}^{n - n_0}\log_2{p_i}'\Big),
\]
\end{enumerate}
where ${p_i}' \in  \{ p_i  \in\Delta_n | p_i > 0 \}$ and $n_0$ is the
number of $p_i$ that $p_i = 0.$
\end{theorem}
\noindent {\it Proof.} We have
\[\begin{array}{lll}
{_\beta}H_n(p_1,p_2,..,p_n) & = (1-\beta)^{-1}\log_2\sum_{i=1}^{n}p_i^\beta  \\ \\
& = (1-\beta)^{-1}\Big(\log_2(n - n_0)\sum_{i=1}^{n - n_0}\frac{{p_i}'^\beta}{n - n_0}\Big) \\ \\
& = (1-\beta)^{-1}\Big(\log_2(n - n_0) +\log_2\sum_{i=1}^{n - n_0}\frac{{p_i}'^\beta}{n - n_0}\Big)\,.\\
\end{array}
\]
Therefore, for $0<\beta<1$, we get
\[
\begin{array}{ll} 
{_\beta}H_n(p_1,p_2,..,p_n) & \geq  
(1-\beta)^{-1}\Big(\log_2(n - n_0) +\sum_{i=1}^{n - n_0}\frac{1}{n - n_0}\log_2{p_i}'^\beta\Big) \\ \\
&= (1-\beta)^{-1}\Big(\log_2(n - n_0) + \frac{\beta}{n - n_0}\sum_{i=1}^{n - n_0}\log_2{p_i}'\Big),
\end{array}
\]
and for $\beta>1$, we get
\[
\begin{array}{ll} 
{_\beta}H_n(p_1,p_2,..,p_n) & \leq 
(1-\beta)^{-1}\Big(\log_2(n - n_0) +\sum_{i=1}^{n - n_0}\frac{1}{n - n_0}\log_2{p_i}'^\beta\Big) \\ \\
&= (1-\beta)^{-1}\Big(\log_2(n - n_0) + \frac{\beta}{n - n_0}\sum_{i=1}^{n - n_0}\log_2{p_i}'\Big)\,.
\end{array}
\]

\qed

From~\eqref{E2} and~\eqref{E3} we have the following relations between the entropy of order $\beta$ and the entropy of type $\beta$ \cite{ZD}: 
\begin{equation}\label{E4}
{_\beta}H_n = (1-\beta)^{-1}\log_2\Big[(2^{1-\beta} - 1)H^\beta_n + 1\Big]
\end{equation}

\begin{theorem}
If $0<\beta<1$, then for all $\textup{(}p_1,p_2,...,p_n\textup{)} \in\Delta_n $ we have 
\[
0 \leq H^\beta_n(p_1,p_2,...,p_n) \leq (2^{1-\beta} - 1)^{-1}\Big[(n - n_0)(\prod_{i=1}^{n - n_0}{p_i}')^\frac{\beta}{n - n_0} - 1\Big],  
\]
where ${p_i}' \in  \{ p_i  \in\Delta_n | p_i > 0 \}$ and $n_0$ is the number of $p_i$ that $p_i = 0.$
\end{theorem}
\noindent {\it Proof.}
For $0<\beta<1$ from~\eqref{E3}, we have:
\[
\begin{aligned}
(1-\beta)^{-1}\log_2\Big[(2^{1-\beta} - 1)H^\beta_n + 1\Big] & \leq  (1-\beta)^{-1}\Big(\log_2(n - n_0) + \frac{\beta}{n - n_0}\sum_{i=1}^{n - n_0}\log_2{p_i}'\Big) \\
\log_2\Big[(2^{1-\beta} - 1)H^\beta_n + 1\Big] & \leq  \log_2\Big[(n - n_0)(\prod_{i=1}^{n - n_0}{p_i}')^\frac{\beta}{n - n_0}\Big],
\end{aligned}
\]
which implies
\[
(2^{1-\beta} - 1)H^\beta_n + 1  \leq (n - n_0)(\prod_{i=1}^{n - n_0}{p_i}')^\frac{\beta}{n - n_0} .\]
Thus,
\[
H^\beta_n(p_1,p_2,...,p_n)  \leq (2^{1-\beta} - 1)^{-1}\Big[(n - n_0)(\prod_{i=1}^{n - n_0}{p_i}')^\frac{\beta}{n - n_0} - 1\Big].
\]
\qed

Product of non-zero probabilities of distribution identifies upper bound, so it is an important criteria for limiting value of entropy of type $\beta$.

\section{Some bounds on the  quantum R{\'e}nyi entropy}
In this section, we're going to extend Theorem \ref{T1} to quantum setting.
In \cite{BSW} the R{\'e}nyi entropy in quantum setting of order $\alpha\in (0,1)\cup (1,\infty)$ is given as
\[H_\alpha (\rho) = \frac{1}{1-\alpha}\log\t{\rho^\alpha}.\]
\begin{theorem}\label{T3}
Let $\rho\in \mathcal{S}$ with $rank(\rho) = d. $ Then, 
\begin{enumerate}
\item if $0<\alpha<1$,
\[
H_\alpha (\rho)\geq \frac{1}{1-\alpha}\Big(\log(d - d_0) + \frac{\alpha}{d - d_0}\log\prod_{i=1}^{d - d_0}{\lambda _i}'\Big). \]
\item if $\alpha>1$,
\[
H_\alpha (\rho)\leq \frac{1}{1-\alpha}\Big(\log(d - d_0) + \frac{\alpha}{d - d_0}\log\prod_{i=1}^{d - d_0}{\lambda _i}'\Big), \]
\end{enumerate}
where ${\lambda _i}' \in  \{ \lambda _i  \in\sigma(\rho) | \lambda _i > 0 \}$ and $d_0$ is the number of $\lambda _i$ that $\lambda _i = 0.$
\end{theorem}
\noindent {\it Proof.}
For any density matrix $\rho$, there exist unitary matrix $U$  and diagonal matrix $D$ such that $D = U\rho U^\dagger$. Also, $D \in \mathcal{S}$. By property of quantum entropy, we have 
\[H_\alpha (D) = H_\alpha (U\rho U^\dagger) = H_\alpha (\rho),\]
\[H_\alpha (D) = \frac{1}{1-\alpha}\log\t{D^\alpha} = \frac{1}{1-\alpha}\log \sum_{i=1}^{d}{\lambda _i}^\alpha,\]
where \[ D = \begin{pmatrix}
\lambda _1 & 0 & \ldots & 0 \\
0 & \lambda _2 & \ldots & 0 \\
\vdots & \vdots & \ddots & \vdots \\
0 & 0  &   \ldots & \lambda _d
\end{pmatrix}\]  and $ \lambda _i  \in\sigma(\rho)$ . 

By Theorem \ref{T1}, we get 
\[
\frac{1}{1-\alpha}\log \sum_{i=1}^{d}{\lambda _i}^\alpha\geq \frac{1}{1-\alpha}\Big(\log(d - d_0) + \frac{\alpha}{d - d_0}\sum_{i=1}^{d - d_0}\log{\lambda _i}'\Big)  \quad \mbox{if} \quad 0<\alpha<1, 
\]
\[
\frac{1}{1-\alpha}\log \sum_{i=1}^{d}{\lambda _i}^\alpha\leq \frac{1}{1-\alpha}\Big(\log(d - d_0) + \frac{\alpha}{d - d_0}\sum_{i=1}^{d - d_0}\log{\lambda _i}'\Big)  \quad \mbox{ if } \quad \alpha>1.
\]  

Thus,
\[
H_\alpha (D)\geq \frac{1}{1-\alpha}\Big(\log(d - d_0) + \frac{\alpha}{d - d_0}\sum_{i=1}^{d - d_0}\log{\lambda _i}'\Big)  \quad \mbox{if} \quad 0<\alpha<1, 
\]
\[
H_\alpha (D)\leq \frac{1}{1-\alpha}\Big(\log(d - d_0) + \frac{\alpha}{d - d_0}\sum_{i=1}^{d - d_0}\log{\lambda _i}'\Big)  \quad \mbox{ if}  \quad \alpha>1.  
\]
Also , $ H_\alpha (D) = H_\alpha (\rho). $ 
\qed

\begin{theorem}
Let $\rho\in \mathcal{S}$ with $rank(\rho) = d. $ Then, 
\[ H_\alpha (\rho) \leq \log d\,. \]
\end{theorem}
\noindent {\it Proof.}
By the convexity of $ t^\alpha (\alpha > 1) $, we obtain 
\[ \sum_{i=1}^{d}{\lambda _i}^\alpha = d(\frac{1}{d}\sum_{i=1}^{d}{\lambda _i}^\alpha) \geq d(\frac{1}{d}\sum_{i=1}^{d}{\lambda _i})^\alpha = d^{1-\alpha}. \]
Thus, for $\alpha > 1 $,
\[ H_\alpha (\rho) = H_\alpha (D) = \frac{1}{1-\alpha}\log \sum_{i=1}^{d}{\lambda _i}^\alpha \leq \frac{1}{1-\alpha}\log d^{1-\alpha} = \log d \]
From concavity of $ t^\alpha (0 <\alpha < 1) $, we obtain
\[ \sum_{i=1}^{d}{\lambda _i}^\alpha = d(\frac{1}{d}\sum_{i=1}^{d}{\lambda _i}^\alpha) \leq d(\frac{1}{d}\sum_{i=1}^{d}{\lambda _i})^\alpha = d^{1-\alpha} \]
Thus, for $0 <\alpha < 1 $,
\[ H_\alpha (\rho) = H_\alpha (D) = \frac{1}{1-\alpha}\log \sum_{i=1}^{d}{\lambda _i}^\alpha \leq \frac{1}{1-\alpha}\log d^{1-\alpha} = \log d.\]
\qed

\begin{corollary}
Let $\rho\in \mathcal{S}$ with $rank(\rho) = d$  and $0<\alpha<1$. Then,
\[ \frac{1}{1-\alpha}\Big(\log(d - d_0) + \frac{\alpha}{d - d_0}\log\prod_{i=1}^{d - d_0}{\lambda _i}'\Big) \leq H_\alpha (\rho)\leq  \log d  \,,\]
where ${\lambda _i}' \in  \{ \lambda _i  \in\sigma(\rho) | \lambda _i > 0 \}$ and $d_0$ is the number of $\lambda _i$ that $\lambda _i = 0.$
\end{corollary}
This theorem provides a lower bound which does not depend on the off-diagonal elements (called as coherences in \cite{DM}), and it also shows that  product of non-zero probabilities of finding the system in the respective states gives us certain least value of quantum R{\'e}nyi entropy.
\section{The R{\'e}nyi relative entropy, conditional entropy, and mutual information}
\begin{definition} \cite{MDOST}
The R{\'e}nyi relative entropy of $ \alpha \geq 0 $ is defined as \\
\[ D_\alpha(\rho || \sigma ) = \frac{1}{\alpha - 1}\log\t\left[\rho^\alpha\sigma^{1 - \alpha}\right]. \]
\end{definition}
\begin{theorem}\label{T4}
For R{\'e}nyi relative entropy of $ \alpha > 1 $, we have  \\
\[ D_\alpha(\rho || \sigma ) \geq \frac{1}{\alpha - 1}\Big(\log d + \frac{\alpha}{d}\log \det(\rho) + \frac{1 - \alpha}{d}\log \det(\sigma)\Big), \]
where $d$ is the dimension of $ \mathcal{H}( \cdot ).$
\end{theorem}
\noindent {\it Proof.}
Since $\rho$ and  $\sigma$ are positive definite matrices, by Lemma \ref{L1},   $ \rho^\alpha$ and $\sigma^{1 - \alpha}$  are also positive definite.
\begin{equation*}
\begin{aligned}
\t(\rho^\alpha \sigma^{1 - \alpha})  & \geq d\Big(\det(\rho^\alpha)\det(\sigma^{1 - \alpha})\Big)^\frac{1}{d} \\
\log \t(\rho^\alpha \sigma^{1 - \alpha}) & \geq \log d + \frac{1}{d}\log \det(\rho^\alpha) + \frac{1}{d}\log \det(\sigma^{1 - \alpha}).
\end{aligned}
\end{equation*}
 Since $ \alpha > 1 $, 
\[ \frac{1}{\alpha - 1 }\log \t(\rho^\alpha \sigma^{1 - \alpha})  \geq \frac{1}{\alpha - 1}\Big(\log d + \frac{1}{d}\log \det(\rho^\alpha) + \frac{1}{d}\log \det(\sigma^{1 - \alpha})\Big). \]
Thus,
\[ D_\alpha(\rho || \sigma ) \geq \frac{1}{\alpha - 1}\Big(\log d + \frac{\alpha}{d}\log \det(\rho) + \frac{1 - \alpha}{d}\log \det(\sigma)\Big). \]
\qed

\begin{definition} \cite{BSW}
From R{\'e}nyi relative entropy, we define R{\'e}nyi generalization of entropy, conditional entropy, and mutual information in analogy respectively with the below formulations;
\begin{equation*}
\begin{aligned}
H_\alpha (A)_\rho & = - D_\alpha (\rho _A || \thinspace \mathbb{I}_A) ,\\
H_\alpha (A|B)_\rho & := \log(dim\mathcal{H}_A) - \min_{\sigma _ B}D_\alpha (\rho _{AB} || \thinspace \mu _A \otimes \sigma _B), \\
I_\alpha (A;B)_\rho & := \min_{\sigma _ B}D_\alpha (\rho _{AB} || \thinspace \rho _A \otimes \sigma _B),
\end{aligned}
\end{equation*}
where  $ \mu _A = \frac{1}{dim\mathcal{H}_A}\mathbb{I}_A.$
\end{definition}
\begin{theorem}\label{T5}
For R{\'e}nyi generalization of conditional entropy and mutual information $ \alpha > 1 $ , if there are $ c_1,c_2 \in \mathbb R^+ $ and $\sigma_B \in \mathcal{H_B}$ that $\mu^{1 - \alpha} _A\otimes \sigma^{1 - \alpha} _B = c_1\rho^\alpha _{AB}$ , $\rho^{1 - \alpha} _A\otimes \sigma^{1 - \alpha} _B = c_2\rho^\alpha _{AB}$, then  we have  \\
\begin{equation*}
\begin{aligned}
H_\alpha (A|B)_\rho & = \log d_A - \frac{1}{\alpha - 1}\Big(\log d_Ad_B + \frac{2\alpha}{d_Ad_B}\log \det(\rho _{AB}) +\log c_1 \Big) \\
 I_\alpha (A;B)_\rho & = \frac{1}{\alpha - 1}\Big(\log d_Ad_B + \frac{2\alpha}{d_Ad_B}\log \det(\rho _{AB}) +\log c_2 \Big) 
\end{aligned}
\end{equation*}
\end{theorem}
\noindent {\it Proof.}
From Theorem \ref{T4}, 
\[ D_\alpha(\rho || \sigma ) \geq \frac{1}{\alpha - 1}\Big(\log d + \frac{\alpha}{d}\log \det(\rho) + \frac{1 - \alpha}{d}\log \det(\sigma)\Big)\,. \]
Equality holds if and only if $\sigma^{1 - \alpha} = c\rho^\alpha $ ; $ c \in \mathbb R^+ $\,. \\
So, if there are $ c_1,c_2 \in \mathbb R^+ $ and $\sigma_B \in \mathcal{H_B}$ that $\mu^{1 - \alpha} _A\otimes \sigma^{1 - \alpha} _B = c_1\rho^\alpha _{AB}$ , $\rho^{1 - \alpha} _A\otimes \sigma^{1 - \alpha} _B = c_2\rho^\alpha _{AB}$ for $ \alpha > 1 $, then  
\begin{equation*}
\begin{aligned}
\min_{\sigma _ B}D_\alpha (\rho _{AB} || \thinspace \mu _A \otimes \sigma _B)  & = \frac{1}{\alpha - 1}\Big(\log d_Ad_B + \frac{\alpha}{d_Ad_B}\log \det(\rho _{AB}) +\frac{1}{d_Ad_B}\log \det(c\rho^\alpha  _{AB})\Big)  \\
& = \frac{1}{\alpha - 1}\Big(\log d_Ad_B + \frac{2\alpha}{d_Ad_B}\log \det(\rho _{AB}) +\log c \Big). 
\end{aligned}
\end{equation*}
Therefore, \\
\[ H_\alpha (A|B)_\rho = \log d_A - \frac{1}{\alpha - 1}\Big(\log d_Ad_B + \frac{2\alpha}{d_Ad_B}\log \det(\rho _{AB}) +\log c_1 \Big). \] 
Similarly, we have these conclusion for $ I_\alpha (A;B)_\rho $ : \\
\[ I_\alpha (A;B)_\rho = \frac{1}{\alpha - 1}\Big(\log d_Ad_B + \frac{2\alpha}{d_Ad_B}\log \det(\rho _{AB}) +\log c_2 \Big), \]
where $ \rho^{1 - \alpha} _A\otimes \sigma^{1 - \alpha} _B = c_2\rho^\alpha _{AB} $, $ c_2 \in \mathbb R^+ $, and $ \alpha > 1 .$ 
\qed

Now, we are going to survey previous theorem in a wider sense. 
\begin{theorem}\label{T6}
For R{\'e}nyi generalization of  mutual information $ \alpha > 1 $, we have  \\
\[ I_\alpha (A;B)_\rho \geq \frac{\alpha}{\alpha - 1}\Big(  \log d_Ad_B +  \frac{1}{d_Ad_B}\log \det(\rho _{AB}) \Big) \]
\end{theorem}
\noindent {\it Proof.}
 \[ \det(\sigma) = \frac{1}{d^d} \det(d \sigma). \]
 From Lemma \ref{L4} and for $ \alpha > 1 $,
\begin{equation*}
\begin{aligned}
\frac{1 - \alpha}{d}\log \det(\sigma) & \geq \frac{1 - \alpha}{d}\Big( \t(d \sigma - \mathbb I) - d\log d \Big) \\
& = (\alpha - 1)\log d
\end{aligned}
\end{equation*}
From Theorem \ref{T4}, 
\begin{equation*}
\begin{aligned}
D_\alpha(\rho || \sigma ) & \geq \frac{1}{\alpha - 1}\Big(\log d + \frac{\alpha}{d}\log \det(\rho) + \frac{1 - \alpha}{d}\log \det(\sigma)\Big)\ \\
& \geq \frac{1}{\alpha - 1}\Big(\log d + (\alpha - 1)\log d + \frac{\alpha}{d}\log \det(\rho)\Big)\ \\
& = \frac{\alpha}{\alpha - 1}\Big( \log d + \frac{1}{d}\log \det(\rho)\Big).\ \\
\end{aligned}
\end{equation*} 
Hence,
\[ I_\alpha (A;B)_\rho \geq \frac{\alpha}{\alpha - 1}\Big(  \log d_Ad_B +  \frac{1}{d_Ad_B}\log \det(\rho _{AB}) \Big) \]
\qed
\\
Therefore, we know the least R{\'e}nyi generalization of  mutual information for $ \alpha > 1 $, before finding optimum $\sigma _B$. \\
For instance, we suppose $ \rho _{AB} = \frac{1}{d_Ad_B}\mathbb I_{AB} $, so 
\[ c_1 = c_2 = (\frac{1}{d_Ad_B})^{1 - 2\alpha} \quad , \quad \rho _A = \t _B(\rho _{AB})  = \frac{1}{d_A}\mathbb I_{A}  \]
and optimum $\sigma _B$ is $\frac{1}{d_B}\mathbb I_{B}$ , we have below conclusions : 
\begin{equation*}
\begin{aligned}
H_\alpha (A|B)_\rho & = \log d_A - \frac{1}{\alpha - 1}\Big(\log d_Ad_B + \frac{2\alpha}{d_Ad_B}\log \det(\frac{1}{d_Ad_B}\mathbb I_{AB}) +\log (\frac{1}{d_Ad_B})^{1 - 2\alpha} \Big) \\
\\
& = \log d_A - \frac{1}{\alpha - 1}\Big(\log d_Ad_B + \frac{2\alpha}{d_Ad_B}\log(\frac{1}{d_Ad_B})^{d_Ad_B} +\log (\frac{1}{d_Ad_B})^{1 - 2\alpha} \Big) \\
\\
& = \log d_A - \frac{1}{\alpha - 1}\Big(\log d_Ad_B + 2\alpha\log(\frac{1}{d_Ad_B}) + (1 - 2\alpha) \log (\frac{1}{d_Ad_B})\Big) \\
\\
& = \log d_A - \frac{1}{\alpha - 1}\Big(\log d_Ad_B + \log(\frac{1}{d_Ad_B})\Big) \\
\\
& = \log d_A = H_\alpha (A)_\rho
\end{aligned}
\end{equation*}
and also, $ I_\alpha (A;B)_\rho = 0.$ 

The last theorem shows relationship between Renyi relative entropy of $\rho$, $\sigma$ and identity for $ \alpha > 1 $.  
\begin{theorem}
For R{\'e}nyi relative entropy of $ \alpha > 1 $, 
\[ D_\alpha(\rho || \sigma ) \leq D_\alpha(\rho || \mathbb{I}) + D_\alpha(\mathbb{I} || \sigma). \]
\end{theorem}
\noindent{\it Proof.}
Using Lemma \ref{L3}, we get 
 \begin{equation*}
\begin{aligned}
\t(\rho^\alpha\sigma^{1 - \alpha}) & \leq \t(\rho^\alpha)\t(\sigma^{1 - \alpha})  \\
\log\t(\rho^\alpha\sigma^{1 - \alpha}) & \leq \log\t(\rho^\alpha) + \log\t(\sigma^{1 - \alpha}) \\
\frac{1}{\alpha - 1}\log\t(\rho^\alpha\sigma^{1 - \alpha}) & \leq \frac{1}{\alpha - 1}\log\t(\rho^\alpha) + \frac{1}{\alpha - 1}\log\t(\sigma^{1 - \alpha}) \quad for \quad  \alpha > 1. 
\end{aligned}
\end{equation*}
So, \\
\[ D_\alpha(\rho || \sigma ) \leq D_\alpha(\rho || \mathbb{I}) + D_\alpha(\mathbb{I} || \sigma). \]
\qed
\bibliographystyle{amsplain}

\end{document}